# Resolving Code Smells in Software Product Line using Refactoring and Reverse Engineering

Sami Ouali

College of Applied Sciences, Ibri, Oman

## ABSTRACT

*Software Product Lines (SPL) are recognized as a successful approach to reuse in software development. Its purpose is to reduce production costs. This approach allows products to be different with respect of particular characteristics and constraints in order to cover different markets. Software Product Line engineering is the production process in product lines. It exploits the commonalities between software products, but also to preserve the ability to vary the functionality between these products. Sometimes, an inappropriate implementation of SPL during this process can conduct to code smells or code anomalies. Code smells are considered as problems in source code which can have an impact on the quality of the derived products of an SPL. The same problem can be present in many derived products from an SPL due to reuse. A possible solution to this problem can be the refactoring which can improve the internal structure of source code without altering external behavior. This paper proposes an approach for building SPL from source code. Its purpose is to reduce code smells in the obtained SPL using refactoring source code. Another part of the approach consists on obtained SPL's design based on reverse engineering.*

## KEYWORDS

*Software Product Line, Code smells, Refactoring, Reverse Engineering.*

## 1. INTRODUCTION

Software Product Line (SPL) is a family of related software systems with common and variable functions whose first goal is reusability [1]. The SPL approach intends at upgrading software productivity and quality by relying on the similarity that exists among software systems, and by managing a family of software systems in a reuse-based way. SPL aims to minimize effort and cost of development and maintenance, to reduce time-to-market and to ameliorate quality of software [2], [3], [4]. Unsuitable development of a SPLs may give rise to bad programming practices, called code anomalies, also referred in the literature as "code smells" [5].

Code smell is often considered as key indicator of something wrong in the system code [5] or undesired code source property. Like all software systems, artifacts of a SPL may contains several code anomalies [6]. Therefore, if these code smells are not systematically removed, the SPL's quality may degrade due to evolution. Code Smells are very-known in classic and single software systems [7]. However, in the context of SPL, Code Smell is a young topic. [8] proposed a specific SPL's smell, called "Variability Smells". [9] discussed two types of bad smells related on SPL: Architectural Bad Smells and Code Bad Smells. [6] and [10] proposed detection strategies for anomalies in SPL.

The main goal of this work is to propose a solution to reduce code smells in SPL. Unsuitable development of a SPLs may give rise to bad practices such as architectural smells and code smells. Our work tries to reduce development problems through the source code analyze of product variants to detect and correct code smells, identify the variability and build the variability model of SPL. Detecting and refactoring code anomalies in source code from the start give us a chance to develop a SPL with a high quality. Thus, the reverse engineering is a preliminary strategy for a clean SPL and to obtain the variability model of SPL.

This paper is organized as follow. Section 2 provides background on code smells, SPL and reverse engineering. Section 3 presents the related work. Section 4 shows the proposed approach. The last section concludes and presents future work.

## 2. BACKGROUND

### 2.1. Software Product Lines

The evolution of software development and the growth of product numbers have motivated the emergence of many reuse concepts. Software development communities recognize SPL as a successful approach for reuse [11], [12]. This success results from the reduction of production costs and time to market. SPL is a software development paradigm that share common feature to satisfy the specific needs of particular market segment [13].

Software product line's approach focus on the sharing of a reference architecture between products. These products can differ and the approach allows this variation with respect of particular characteristics and constraints. This difference is the variability present in SPL, which is the ability of a core asset to adapt to usages in the different product contexts that are within the product line scope [14]. Variability must be anticipated and continuously maintained to obtain wished results. The production process of product lines is well known as software product line engineering (SPLE) which tries to maximize the commonalities and reduce the cost of variations [15]. The SPLE process focuses on two levels of engineering [14]: Domain Engineering (DE) and Application Engineering (AE). DE focuses on developing reusable artifacts which are used in AE to construct a specific product. Fig. 1 presents the SPLE process.

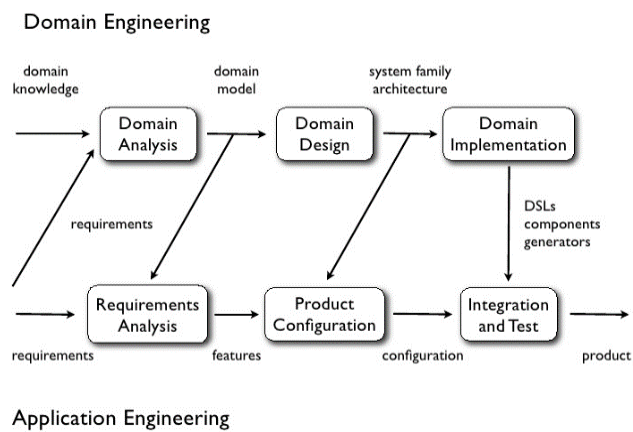

Figure 1.  Domain Engineering and Application Engineering [14]

## 2.2. Code smells

A software system evolves over time. Its evolution is one of the critical phases of the process of its development. Moreover, the software system changes, moreover the structure of the program deteriorates. So, complexity increases until it becomes more profitable to rewrite it from the scratch. Which can involve threats on the software quality.

Software system's bad quality is a key indicator of existing bad programming practices, also known in the literature as source code flaw, code smells or code anomalies [5].

Code smells are usually symptoms of low-level problems such as anti-patterns. They are indicators of something wrong that structures in the source code [5], their presence can affect in maintenance and slow down software development.

In literature, different Code Smells have been defined. For instance, in Fowler's book [5], Beck define a list of 22 code smells, for example "Long Method" is a method that is too long and has too many responsibilities, so it makes code hard to maintain, understand, change, extend, debug and reuse. "Large Class" is a class contains many fields, methods or lines of code, means that a class is trying to do too much. "Duplicated Code" has negative impacts on software development and maintenance. For example, they increase bug occurrences: if an instance of duplicate code is changed in one part of the code for fixing bugs or adding new features, code may require various changes in other parts all over the source code simultaneously; if the correspondents are not changed inadvertently, bugs are newly introduced to them [16].

## 2.3. Reverse Engineering

Reverse Engineering is the process of analyzing a system. The purpose is to identify system structure, its components and the relationships between them [17].

Reverse Engineering can create representations of the system through transformations between or within different abstraction levels. It can also extract design information from source code [17] and may be used to re-implement the system.

The reverse engineering process can be done through automated analysis or manual annotations. The next steps concern the identification of program structure and the establishment of traceability matrix.

## 2.4. Refactoring

Refactoring's purpose is to improve the quality of an existing code [5]. This process tries through the software system changing to improve its internal structure without having an impact on the external behavior of the code.

Refactoring can be a solution for code smells. This process takes as input a source code with problems and outputs good ones. The resulting code can be reused. The refactoring allows the code smells identification. Also, it offers the possibility to change the original code containing these code smells by code restructuration to get an output code without code smells.

## 3. RELATED WORK

Common industrial practices lead to the development of similar software products, then they are usually managed to each other using simple techniques, e.g., copy-paste-modify. This is bad practice leading a low software quality, as we mentioned above the "Duplicated Code" code smell. During the past few years, several studies have investigated two things: how to detect code smells [18], [19], [20], [21], [22], [23] and how to correct [5], [18], [24] them in a single software. To the best of our knowledge we found few studies [6], [8], [9], [10], [25], [26] that can be considered related to our research.

[9] performed a Systematic Literature Review (SLR) to find and classify published work about bad smells in the context of SPL and their respective refactoring methods. They classified 70 different bad smells divided in three groups: (i) Code Smells; that are symptoms of something wrong in the source code, (ii) Architectural Smells; that are an indication of problem in higher levels of abstraction and (iii) hybrid Smells; that are a combination between architectural smell and code smells. [26] proposed a method to derive metric thresholds for software product lines.
The goal is to define thresholds values that each metric can take in order to identify potential problems in the implementation of features. They use 4 software metrics: Lines of Code (LOC) counts the number of uncommented lines of code per class. The value of this metric indicates the size of a class. Coupling between Objects (CBO) counts the number of classes called by a given class. CBO measures the degree of coupling among classes. Weight Method per Class (WMC) counts the number of methods in a class. This metric can be used to estimate the complexity of a class. Number of Constant Refinements (NCR) counts the number of refinements that a constant has. Its value indicates how complex the relationship between a constant and its features is. Their study is based on 33 SPLs which are divided into three benchmarks according to their size in terms of Lines of Code (LOC).

Benchmark 1 includes all 33 SPLs. Benchmark 2 includes 22 SPLs with more than 300 LOC. Finally, Benchmark 3 is composed of 14 SPLs with more than 1,000 LOC. The goal of creating three different benchmarks is to analyze the results with varying levels of thresholds. In term of that they illustrate a detection strategy to detect two types of code smells, God Class and Lazy Class. Figure 2 presents the way to identify God Class and Lazy Class.

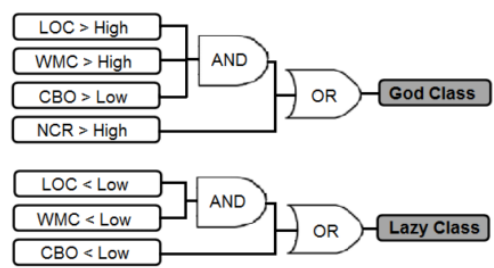

Figure 2. Code Smells identification.

Apel et al. [8] proposed bad smell specific to SPLs called variability smell; that is an indicator of an existing undesired property in all kinds of artifacts in an SPL, such as feature models.

Fernandes and Figueiredo [6] investigated code anomalies in the context of SPLs, they propose new detection strategies for well-known anomalies in SPL such as God Class and God Method, ultimately they propose new anomalies and their detection strategies and they propose supporting tool for the proposed detection.

De Andrade et al. [25] conducted an exploratory study that aims at characterizing architectural smells in the context of software product line.

Abilio et al. [10] proposed means to detect three code smells (God Method, God Class, and Shotgun Surgery) in Feature-Oriented Programming source code, FOP is a specific technique to deal with the modularization of features in SPL. They performed an exploratory study with eight SPLs developed with AHEAD; which is an FOP language, to detect code smells in a SPL by using 16 source code metrics. These metrics corresponds to the detection of three code smells mentioned above. Table 1 presents some of these metrics.

Table 1.  Metrics used to detect code smells [10]

| Acronym | Name | Description |
| --- | --- | --- |
| NOF | Number of Features | Number of Features which has code artifacts |
| NCR | Number of Constant Refinements | Number of refinements which a constant has |
| NMR | Number of Method Refinements | Number of refinements which a method has |
| TNCt | Total Number of Constants | Number of constants (classes, interfaces - constant) |
| TNR | Total Number of Refinements | Total of refinements (classes, interfaces - refinement) |
| TNMR | Total Number of Method Refinements | Total of refinements of a method |
| TNRC | Total Number of Refined Constants | Total of refined constants |
| TNRM | Total Number of Refined Methods | Total of refined methods |

Considering the discussed related work, we propose an approach aiming to develop an SPL with minimal code smells risks.

## 4. PROPOSED APPROACH

The main goals in our study are to (i) investigate the state of the art on code smells in the context of SPLs as we show above, (ii) propose a solution to decrease code smells in developing software product lines.

Unsuitable development of a SPLs may give rise to bad practices such as architectural smells and code smells that induce maintenance and development costs problems. Therefore, we propose to build an SPL from the scratch using reverse engineering methods, which can help us to detect and correct code smells from the start. Thus, we can guarantee great quality of SPL.

The main challenge in this task is to analyze the source code of product variants in order to (i) detect and correct code smells, (ii) identify the variability among the products, (iii) associate them with features and (iiii) regroup the features into a variability model. The proposed approach is object-oriented language and only uses as input the source code of product variants.

First of all, we use as input source code of product variants then we apply detection strategies for code anomalies as duplicated code, uncovered code by unit tests and too complex code, after that we correct them using an automated bad smell correction technique based on the generation of refactoring concepts. Refactoring is a change made to the internal structure of software to rewrite the code, to "clean it up", to make it easier to understand and cheaper to modify without changing its observable behavior [27]. In step 2 and after having a clean code, we are interested in the determination of the semantic relations between the names of the classes, the names of the methods and the attributes of all the source codes of the existing products having different terminologies and not necessary having the same meaning. In term of that we are interested in the harmonization of names, and more particularly in unifying fragments of source codes. During unification, we determine the semantic correspondences between the source code elements based on semantic knowledge base YAGO [28].

YAGO is a semantic knowledge base derived from many data sources like Wikipedia, WordNet, WikiData, GeoNames, and other. Aside YAGO, we will base on Machine Learning methods to get better semantic correspondences between source code elements. In fact, Machine Learning algorithms can be helpful in the classification of the features. Machine Learning proved his efficiency in many complex domains like Predictive Analytics [29], image processing [30], and signal processing… At the end of this step, all names with a semantic relationship would be harmonized and can be further analyzed in the next step of identifying commonalities and variability. Thus, we extract features by identification of common block (CB) and variation blocks (VB). CB groups the elements present in all the products while VB groups the elements present in certain products and not all of them. The role of these blocks is to group subsets to implement features. Once the common block and the variation blocks are completed, the extraction of mandatory elements and variation atomic blocks is supported, we associate them to features. Once the common properties and variability of product variants are identified, the feature model(s) will be constructed. Consequently, we can obtain one or more than one SPL. Our approach is presented in Figure3.

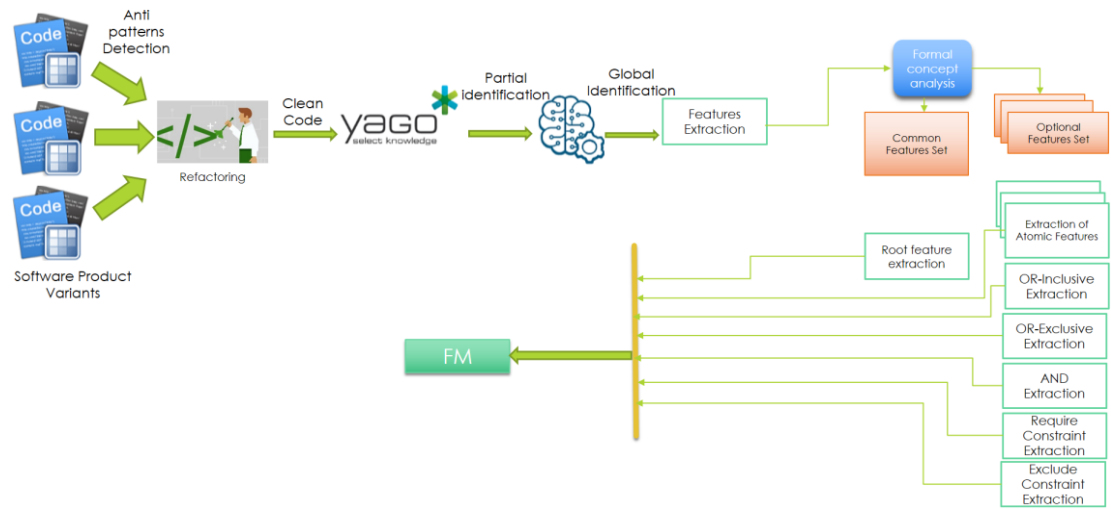

Figure 3. Proposed Approach.

## 5. CONCLUSIONS

Software reuse is an important challenge in software engineering. Software Product Line is one of the technique used to ensure the success of this challenge. The obtained products can contain reused parts or components. These parts can include some problems in their source code more known as Code Smells. These problems can propagate between the different products.

A solution to avoid the Code smells in source code, is refactoring which can improve the internal structure of software system by trying to find the problem and avoid it using some restructuration techniques.

In this paper, we try to present an approach which combines refactoring to eliminate code smells and reverse engineering to propagate modifications to the design level. Our purpose is to obtain a software product line model free from code smells.

Our future works will be the refinement of the different parts of the approach. Also, we will choose the appropriate tools to use in our prototype.

## AUTHORS

**Sami Ouali** is an assistant professor in the College of Applied Sciences of lbri in Oman. He is a member of the RIADI labs, Tunisia. His research interests lie in the areas of software engineering and software product line.

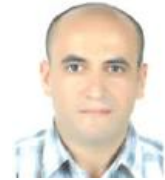